\documentclass[epj]{svjour} 
\usepackage{graphicx}
\begin{document}
\titlerunning{On the Controversy around Second-Order Traffic Flow Models}
\title{On the Controversy around Daganzo's Requiem for and Aw-Rascle's Resurrection of Second-Order Traffic Flow Models}
\author{Dirk Helbing and Anders Johansson
}                     
%
%
\institute{ETH Zurich, UNO D11, Universit\"atstr. 41, 8092 Zurich, Switzerland}
\date{Received: date / Revised version: date}
%
\abstract{Daganzo's criticisms of second-order fluid approximations of traffic flow [C. Daganzo, Transpn. Res. B. 29, 277-286 (1995)] and Aw and Rascle's proposal how to overcome them [A. Aw and M. Rascle, SIAM J. Appl. Math. 60, 916-938 (2000)] have stimulated an intensive scientific activity in the field of traffic modeling. Here, we will revisit their arguments and the interpretations behind them. We will start by analyzing the linear stability of traffic models, which  is a widely established approach to study the ability of traffic models to describe emergent traffic jams. Besides deriving a collection of useful formulas for stability analyses, the main attention is put on the characteristic speeds, which are related to the group velocities of the linearized model equations. Most macroscopic traffic models with a dynamic velocity equation appear to predict {\it two} characteristic speeds, one of which is {\it faster} than the average velocity. This has been claimed to constitute a theoretical inconsistency. We will carefully discuss arguments for and against this view. In particular, we will shed some new light on the problem by comparing Payne's macroscopic traffic model with the Aw-Rascle model and macroscopic with microscopic traffic models. 
} 
\maketitle
\section{Introduction}

Understanding traffic congestion has puzzled not only traffic engineers, but also a large number of physicists \cite{Schadschneider,Review,Nagatani,Nagel}. Scientists have been particularly interested in emergent traffic jams, which are related to instabilities in the traffic flow. Such instabilities have been found in empirical data \cite{TranSci}, but also in recent experiments \cite{Sugiyama}.
\par
The theoretical analysis is usually done by computer simulation or by linear stability analysis.
Both techniques have been used since the early days of traffic engineering \cite{Herman} and traffic physics \cite{Kuehne,Bando}. Here, we will perform the analysis for macroscopic {\it and} microscopic models {\it in parallel}, as there should be a correspondence between the properties of both kinds of models. In contrast to previous publications, the analysis of macroscopic traffic equations is done for a model that considers a dependence of the optimal velocity function and the traffic pressure on the average velocity, not only the density. Such a dependence results for models which represent vehicle interactions realistically, taking into account a velocity-dependent safety distance \cite{EPJB}. This is, for example, important to avoid accidents, and it changes the instability conditions significantly (see Sec. \ref{Macinstab}).
\par
Besides determining the stability threshold, a particular focus will be put on the calculation of the group velocities of the partial differential equations underlying the macroscopic traffic model (see Sec. \ref{Macspeeds}). For clarity, the definition of the group velocities will be compared with those of phase velocities and of characteristic speeds. All three definitions describe propagation processes of waves. It will be shown, that they lead to identical results under certain circumstances, but not necessarily so. 
\par
Furthermore, we will derive conditions under which one of the group velocities is greater than the average velocity. In Sec. \ref{Hist}, we will shortly summarize the main points of the controversial discussion that this observation has triggered. We will also address Daganzo's other criticisms of second-order fluid approximations of traffic flow \cite{Daganzo}. After the formal analysis in Sec. \ref{Macinstab}, Sec. \ref{Discuss} will be dedicated to a careful discussion of the results. {In particular, we will analyze different conceivable reasons for characteristic speeds faster than the vehicle speeds: (1) artifacts due to approximations underlying second-order macroscopic traffic models, (2) indirect long-range {\it forward} interactions with {\it followers} on a circular road, (3) the definition of the propagation speed of perturbations, (4) the variability of vehicle velocities, (5) the interpretation of characteristic speeds. 
Since characteristic speeds} are primarily perceived as a problem of second-order macroscopic traffic models,  in Sec. \ref{Micinstab} we will compare them with the group velocities predicted by microscopic traffic models. Finally, we will summarize our results in Sec. \ref{Conclusions}.

\section{Summary of the Controversy regarding Second-Order Traffic Flow Models}\label{Hist}

In the area of macroscopic traffic flow modeling, it is common to formulate equations for the vehicle density $\rho(x,t)$ as a function of space $x$ and time $t$ and for the average velocity $V(x,t)$. The most well-known model, sometimes called the LWR model, was proposed by Lighthill, Whitham, and Richards \cite{LiWi,Ri}. It is based on the continuity equation 
\begin{equation}
 \frac{\partial \rho(x,t)}{\partial t} + V(x,t) \frac{\partial \rho(x,t)}{\partial x} 
 = - \rho(x,t) \frac{\partial V(x,t)}{\partial x}
 \label{g1}
\end{equation}
for the density and a speed-density relationship
\begin{equation}
V(x,t) = V_{\rm e}\big(\rho(x,t)\big)
\end{equation}
or, alternatively, a ``fundamental diagram'' $Q(x,t) = Q_{\rm e}(\rho(x,t))$ for the vehicle flow $Q(x,t)=\rho(x,t)V(x,t)$. Obviously, the LWR model is based on a (hyperbolic) partial differential equation of first order. A detailed analysis is given in Refs. \cite{LiWi,Whitham}. It is well-known, that it describes the generation of shock waves characterized by discontinuous density changes. 
\par
Therefore, in his famous ``Requiem for Second-Order Fluid Approximations of Traffic Flow'' \cite{Daganzo}, Carlos Daganzo correctly notes on page 285 that, ``Besides a coarse representation of shocks, other deficiencies of the LWR theory include its failure to describe {\it platoon diffusion} properly ... and its inability to explain the instability of heavy traffic, which exhibits oscillatory phenomena on the order of minutes.'' However, he also criticizes theoretical inconsistencies of alternative models, which, at that time, were mainly second-order models containing diffusion, pressure, or viscosity terms. The Payne-Whitham model \cite{Whitham,Payne1,Payne2}, for example, has a dynamic velocity equation of the form
\begin{eqnarray}
& & \frac{\partial V(x,t)}{\partial t} + V(x,t)\frac{\partial V(x,t)}{\partial x} \nonumber \\
&=& - \frac{\nu}{\rho(x,t)}\frac{\partial \rho(x,t)}{\partial x} + \frac{1}{\tau} \Big[ V_{\rm e}\big(\rho(x,t)\big) - V(x,t)\Big] \label{Pa1}
\end{eqnarray}
with
\begin{equation}
\nu = - \frac{1}{2\tau} \frac{dV_{\rm e}(\rho)}{d\rho} = 
\frac{1}{2\tau} \left| \frac{dV_{\rm e}(\rho)}{d\rho}\right|  \ge 0 \, . \label{Pa2}
\end{equation}
Here, the term containing $\nu$ is called {\it anticipation term}, while the last term is known as {\it relaxation term}. $V_{\rm e}(\rho)$ denotes the equilibrium velocity and $\tau$ the relaxation time.
\par
Some of the second-order models, including the Payne-Whitham model \cite{Payne1,Payne2}, can be derived from car-following models by certain approximations. This involves gradient expansions of non-local, forwardly directed (i.e. anisotropic) vehicle interactions \cite{EPJB}. Such approximations are problematic, since they lead to terms containing spatial derivatives, which imply undesired backward interaction effects as well. The related theoretical inconsistencies were elaborated by Daganzo. In the following, we will summarize his critique by quotes from Ref. \cite{Daganzo} (page numbers in square brackets):
\begin{enumerate}
\item {\bf Lack of anisotropy:} ``A fluid particle responds to stimuli from the front and from behind, but a car is an anisotropic particle that mostly responds to frontal stimuli'' [p. 279].
\item {\bf Insufficient description of jam fronts:} ``The width of a traffic shock only encompasses a few vehicles'', while second-order models involving viscosity terms would typically imply extended jam fronts [p. 279]. Daganzo argues that ``the smoothness of the shock is inherently unreasonable'' [p. 282], because ``spacings and density must change abruptly whenever the road behind is empty'' [p. 282]. Based on the analysis of concrete examples, Daganzo further finds that ``the cars at the end of the queue move back and the behavior spreads to the remaining vehicles in the queue ... from the back to the front!'' [p. 283]. Further on, new arrivals of vehicles would ``compress a queue from behind'' [p. 283]. 
\item {\bf Insufficient representation of acceleration processes and driver characteristics:} According to the ``relaxation''  mechanism for the velocity distribution assumed in the gas-kinetic traffic model by Prigogine {\it et al.} \cite{Prigogine}, the ``desired speed distribution is a property of the road and not the drivers, as noted by Paveri-Fontana (1975)'' [p. 280]. However, ``Unlike molecules, vehicles have personalities (e.g., aggressive and timid) that remain unchanged by motion'' [p. 279], and models should make sure ``that interactions do not change the `personality' (agressive/timid) of any car'' [p. 280]. Therefore, ``a slow car should be virtually unaffected by its interaction with faster cars passing it (or queueing behind it) ...'' [p. 280].
\end{enumerate}
A further criticism concerns the propagation speeds of perturbations in the traffic flow, predicted by second-order traffic models, which will be addressed after we have replied to the above, well-taken points:
\begin{enumerate}
\item The lack of anisotropy is a consequence of gradient expansions and can be avoided by non-local macroscopic traffic models \cite{EPJB}, such as the gas-kinetic-based traffic model (GKT model) \cite{GKT,GKTmulti}.
\item Non-local traffic models can represent sharp shock fronts well, as has been demonstrated for the GKT model \cite{numerics}. They are also capable of avoiding negative vehicle velocities, if properly specified \cite{numerics}. For example, the speed variance $\theta$ appearing in some macroscopic traffic models, in particular in the ``pressure term'' (see below) must vanish, whenever the average velocity $V$ vanishes. This can be reached by a relationship of the form $\theta(\rho,V) = \alpha(\rho)V^2$ with a suitable, density-dependent function $\alpha(\rho) \ge 0$ \cite{GKT,GKTmulti}.
\item The personality of drivers can be represented by multi-class traffic models \cite{GKTmulti,Wagner,Hoog}. Moreover,
the unrealistic acceleration-behavior implied by Prigogine's gas-kinetic traffic model \cite{Prigogine} has been overcome by the gas-kinetic model by Paveri-Fontana \cite{Paveri} and its generalizations to different driver-vehicle classes \cite{GKTmulti,Wagner}. In these models, it is {\it not} the velocity {\it distribution} which relaxes to a {\it desired} velocity distribution (which would imply discontinuous velocity jumps at a certain rate). Rather they describe a continuous adaptation of individual vehicle velocities to their desired speeds.
\end{enumerate}
Let us now turn to the discussion of the ``characteristic speeds''. Characteristic speeds relate to the eigenvalues of hyperbolic partial differential equations. They determine the solutions for given initial and boundary conditions, in particular which locations influence the solution at other locations at a given time \cite{PartD1,PartD2} (see Appendix \ref{HPDE}). The characteristic speeds are also 
important for the stability of numerical solution schemes for partial differential equations \cite{PartD3}.
\par
What implications does this have for macroscopic traffic models based on systems of hyperbolic partial differential equations with source terms? In his ``Requiem for second-order fluid approximations of traffic flow'' \cite{Daganzo}, Daganzo argues that ``high-order models always exhibit one characteristic speed greater than the macroscopic fluid velocity. ... This is highly undesirable because it means that the future conditions of a traffic element are, in part, determined by what is happening ... BEHIND IT! ... it is a manifestation of the erroneous cause and effect relationship between current and future variables that is at the heart of all high-order models'' [p. 281]. 
\par
{Is this violation of causality a result of crude approximations underlying second-order macroscopic traffic models? Or could the assumption of circular boundary conditions explain an influence from behind, even in the case where vehicle interactions are exclusively directed to the front? Or is the faster characteristic speed related to vehicle interactions at all?}
Until today, the problem of characteristic speeds is puzzling, and it has stimulated many scientists to develop and investigate improved macroscopic traffic 
models \cite{Rascle,Klar,Green,Zhang,Goatin,Piccoli,Siebel,Dai,Lebacque,Degond}. 
Here, we restrict our discussion to the most prominent example: In their ``Resurrection of `second order' models of traffic flow'' \cite{Rascle}, Aw and Rascle propose a new model with two characteristic speeds, one of which is smaller than and the other one equal to $V$, where $V$ denotes the macroscopic vehicle speed. Details are discussed in Sec. \ref{AR}. While, without any doubt, such an approach is interesting and worth pursuing, we will address the question, whether it is {\it necessary} to overcome the problem pointed out by Daganzo. This issue must be analyzed very carefully in order to exclude misunderstandings and to avoid jumping to a conclusion. To provide a complete chain of arguments, the main text of this paper is supplemented by several appendices.

\section{Linear Instability of Macroscopic Traffic Models} \label{Macinstab}

Let us start our analysis with the continuity equation (\ref{g1}) for the vehicle density $\rho(x,t)$
and a macroscopic equation for the average velocity $V(x,t)$ of the type derived at the end of Sec. 4.4.3
of Ref. \cite{EPJB}: Assuming repulsive vehicle interactions that depend on the vehicle distance and vehicle speed, but (for simplicity) not on the relative velocity, it reads
\begin{eqnarray}
& & \frac{\partial V(x,t)}{\partial t} + V(x,t) \frac{\partial V(x,t)}{\partial x} \nonumber \\ 
&=& - \frac{1}{\rho}\frac{\partial P_1(\rho,V)}{\partial \rho} \frac{\partial \rho(x,t)}{\partial x} 
- \frac{1}{\rho}\frac{\partial P_2(\rho,V)}{\partial V} \frac{\partial V(x,t)}{\partial x} \nonumber \\
&+& \frac{V_{\rm o}(\rho,V) - V(x,t)}{\tau} \, .
\label{veleq}
\end{eqnarray}
Herein, $P_1$ and $P_2$ are contributions to the ``traffic pressure'', 
and $V_{\rm o}(\rho,V)$ is the ``optimal velocity'' function. 
\par
Our stability analysis starts with an initial state of uniform vehicle density $\rho_{\rm e}$. The related stationary and homogeneous (i.e. time- and location-independent) solution is obtained by setting the partial derivatives $\partial/\partial t$ and $\partial/\partial x$ to zero. In this way, 
Eq. (\ref{veleq}) yields the implicit equation
\begin{equation}
V_{\rm e}(\rho_{\rm e}) = V_{\rm o}\big(\rho_{\rm e},V_{\rm e}(\rho_{\rm e})\big) 
\end{equation}
for the equilibrium speed  $V_{\rm e}(\rho_{\rm e})$. With this, we can define the deviations
\begin{equation}
 \delta \rho(x,t) = \rho(x,t) - \rho_{\rm e} \qquad \mbox{and} \qquad \delta V(x,t) = V(x,t) - V_{\rm e} \, .
 \label{Devia}
\end{equation}
Inserting $\rho(x,t) = \rho_{\rm e} + \delta \rho(x,t)$ and $V(x,t) = V_{\rm e} + \delta V(x,t)$ into the continuity equation, performing Taylor approximations, where necessary, and dropping all non-linear terms because of the assumption
of small deviations $\delta \rho(x,t)/\rho_{\rm e}$ $\ll 1$ and $\delta V(x,t)/V_{\rm e}\ll 1$, we end up with the following
linearized equation:
\begin{equation}
 \frac{\partial \, \delta \rho(x,t)}{\partial t}
 + V_{\rm e}(\rho_{\rm e}) \frac{\partial \, \delta \rho(x,t)}{\partial x}
 = - \rho_{\rm e} \frac{\partial \, \delta V(x,t)}{\partial x} \, .
\label{linrho}
\end{equation}
Analogously, the linerarized dynamical equation for the average velocity becomes
\begin{eqnarray}
 & & \frac{\partial \, \delta V(x,t)}{\partial t}
 + V_{\rm e} \frac{\partial \, \delta V(x,t)}{\partial x} \nonumber \\
 &=& - \frac{1}{\rho_{\rm e}} \bigg[ \frac{\partial {P}_1(\rho_{\rm e},V_{\rm e})}
 {\partial \rho} \frac{\partial \, \delta \rho(x,t)}{\partial x} 
 + \frac{\partial {P}_2(\rho_{\rm e},V_{\rm e})}
 {\partial V} \frac{\partial \, \delta V(x,t)}{\partial x} \bigg] \nonumber \\
 &+& \frac{1}{\tau} \bigg[ \frac{\partial V_{\rm o}(\rho_{\rm e},V_{\rm e})}
 {\partial \rho} \, \delta \rho(x,t) \nonumber \\
 & & + \frac{\partial V_{\rm o}(\rho_{\rm e},V_{\rm e})}
 {\partial V} \, \delta V(x,t)  - \delta V(x,t) \bigg] .
\label{linvau3}
\end{eqnarray}
The terms on the right-hand side in the first square bracket may be considered to describe dispersion and interaction effects contributing to the ``traffic pressure'', while the terms in the second square bracket result from the so-called relaxation term, i.e. the adaptation of the average velocity $V(x,t)$ to some ``optimal velocity'' $V_{\rm o}(\rho,V)$ with a relaxation time $\tau$.
\par
As is shown in Appendix \ref{MAC}, a linear stability analysis of Eqs. (\ref{linrho}) and (\ref{linvau3}) 
leads to the characteristic polynomial
\begin{eqnarray}
 (\tilde{\lambda})^2 &+& \tilde{\lambda} 
 \left[ \frac{{\rm i}\kappa}{\rho_{\rm e}} \frac{\partial P_2}{\partial V}
 + \frac{1}{\tau}\left( 1 - \frac{\partial V_{\rm o}}{\partial V} \right) \right] \nonumber \\
 &+& {\rm i}\kappa\rho_{\rm e} \left( - \frac{{\rm i}\kappa}{\rho_{\rm e}}
 \frac{\partial P_1}{\partial \rho} + \frac{1}{\tau}
 \frac{\partial V_{\rm o}}{\partial \rho} \right) = 0 \, .  
\label{monopol3}
\end{eqnarray}
It has the two solutions (eigenvalues) 
\begin{eqnarray}
\tilde{\lambda}_\pm(\rho_{\rm e},\kappa)  &=& \lambda_{\pm}(\rho_{\rm e},\kappa) - {\rm i} \tilde{\omega}_{\pm}(\rho_{\rm e},\kappa) \nonumber \\
 &=&  - \frac{1}{2\hat{\tau}} - \frac{{\rm i}\kappa}{2\rho_{\rm e}} \frac{\partial P_2}{\partial V} \pm \sqrt{ \Re \pm {\rm i} |\Im| } \qquad 
 \label{lab0}
\end{eqnarray}
with
\begin{eqnarray}
 \frac{1}{\hat{\tau}(\rho_{\rm e},\kappa)} &=& \frac{1}{\tau} \left( 1 - \frac{\partial V_{\rm o}}{\partial V} \right) 
 \ge 0 \, , \label{defi1} \\
\Re(\rho_{\rm e},\kappa) &=& \frac{1}{4\hat{\tau}^2} - \kappa^2 \frac{\partial  P_1}{\partial \rho} 
- \frac{\kappa^2}{4\rho_{\rm e}{}^2} \left( \frac{\partial P_2}{\partial V}\right)^2 \, , \label{defi2} \\ 
 \pm |\Im(\rho_{\rm e},\kappa)|  &=& - \frac{\kappa \rho_{\rm e}}{\tau} \frac{dV_{\rm o}}{d\rho}  
  + \frac{\kappa}{2\rho_{\rm e} \hat{\tau}} \frac{\partial P_2}{\partial V} \, . \label{defi3}
\end{eqnarray}
Here, we have used the abbreviations
\begin{equation}
 \tilde{\lambda} = \lambda - {\rm i} \tilde{\omega} \qquad \mbox{and} \qquad
 \tilde{\omega} = \omega - \kappa  V_{\rm e}(\rho_{\rm e}) \, . 
\label{tilde}
\end{equation}
\par
As the square root contains a complex number, it is difficult to see
the sign of the real value $\lambda$ of $\tilde{\lambda}$. However, we may apply the formula
\begin{eqnarray}
 \sqrt{\Re \pm {\rm i}|\Im |} &=& \sqrt{\frac{1}{2} \Big( \sqrt{\Re^2 + \Im^2} + \Re \Big) } \nonumber \\
 &\pm & {\rm i}  \sqrt{\frac{1}{2} \Big( \sqrt{\Re^2 + \Im^2} - \Re \Big) } \, ,
\label{magic}
\end{eqnarray}
which is derived in Appendix \ref{Magic}. 
From this and Eq. (\ref{lab0}), we get the following relationship for the real part of the eigenvalues $\tilde{\lambda}_\pm(\rho_{\rm e},\kappa)$:
\begin{equation}
 \lambda_\pm(\rho_{\rm e},\kappa )  = \mbox{Re} \big(\tilde{\lambda}_\pm(\rho_{\rm e},\kappa )\big) = - \frac{1}{2\hat{\tau}} \pm \sqrt{\frac{1}{2} \!
 \left(\! \sqrt{\Re^2 + \Im^2} + \Re \right) } .
 \label{real}
\end{equation}
The expression for the imaginary part gives
\begin{eqnarray}
 - \tilde{\omega}_\pm(\rho_{\rm e},\kappa )  &=& \mbox{Im} \big(\tilde{\lambda}_\pm(\rho_{\rm e},\kappa )\big) \nonumber \\
&=& - \frac{\kappa}{2\rho_{\rm e}} \frac{\partial P_2}{\partial V}  
 \pm \sqrt{\frac{1}{2} \left( \sqrt{\Re^2 + \Im^2} - \Re \right) } \, .\qquad
\label{imag}
\end{eqnarray}

\subsection{Derivation of the Instability Condition}

A transition from stable to unstable behavior, i.e. the change from negative to positive
values of $\lambda_\pm(\rho_{\rm e},\kappa )$ occurs only for the 
eigenvalue $\tilde{\lambda}_{_+}(\rho_{\rm e},\kappa)$, namely under the condition
\begin{equation}
 \lambda_+(\rho_{\rm e},\kappa ) = - \frac{1}{2\hat{\tau}} + \sqrt{\frac{1}{2}
 \left( \sqrt{\Re^2 + \Im^2} + \Re \right) } = 0 \, .
\end{equation}
This implies 
\begin{equation}
 \left( \frac{1}{4\hat{\tau}^2} - \frac{\Re}{2} \right)^2  = \frac{1}{4} (\Re^2 + \Im^2 ) 
\end{equation}
and, therefore,
\begin{equation}
 \frac{1}{16\hat{\tau}^4} = \frac{\Re}{4\hat{\tau}^2} + \frac{\Im^2}{4} \, .
 \label{sechzehn}
\end{equation}
Inserting the above definitions of $\Im$ and $\Re$, we eventually find
\begin{eqnarray}
& & \frac{\kappa^2}{4\hat{\tau}^2} \left[  \frac{\partial P_1}{\partial \rho} + \frac{1}{4\rho_{\rm e}{}^2}
 \left( \frac{\partial P_2}{\partial V} \right)^2\right] \nonumber \\
&=& \frac{1}{4} \left( - \frac{\kappa\rho_{\rm e}}{\tau} \frac{\partial V_{\rm o}}{\partial \rho}
  + \frac{\kappa}{2\rho_{\rm e} \hat{\tau}} \frac{\partial P_2} {\partial V} \right)^2 \, .
\end{eqnarray}
From this and definition (\ref{defi1}), we can derive the following condition for the instability threshold:
\begin{equation}
\frac{1}{\hat{\tau}}
\sqrt{ \frac{\partial P_1}{\partial \rho} + \frac{1}{4\rho_{\rm e}{}^2} 
 \left( \frac{\partial P_2}{\partial V} \right)^2} 
= - \frac{\rho_{\rm e}}{\tau} \frac{\partial V_{\rm o}}{\partial \rho}
  + \frac{1}{2\rho_{\rm e} \hat{\tau}} \frac{\partial P_2} {\partial V} \, .
\end{equation}
Assuming the relationships $\partial V_{\rm o}(\rho)/\partial \rho \le 0$, 
$\partial V_{\rm o}/\partial V \le 0$, and $\partial P_2/\partial V \le 0$,
the condition for $\mbox{Re}(\tilde{\lambda}_{_+}) > 0$  becomes
\begin{eqnarray}
\rho_{\rm e} \left| \frac{\partial V_{\rm o}}{\partial \rho}\right|
 &>& \left[ \sqrt{ \frac{\partial P_1}{\partial \rho} 
 +  \frac{1}{4\rho_{\rm e}{}^2} \left( \frac{\partial P_2}{\partial V} \right)^2} 
 + \frac{1}{2\rho_{\rm e}}\left|
 \frac{\partial P_2}{\partial V}\right| \right] \nonumber \\
 &\times &  \left( 1 + \left|\frac{\partial V_{\rm o}}{\partial V}\right| \right) \, .  
 \label{conseq}
\end{eqnarray}
We notice that this instability condition is {\it not} fulfilled, if the
average velocity $V_{\rm o}(\rho,V)$ changes little with the density $\rho$, which
is typically the case for small densities and, in many models, also for large ones. 
However, $\lambda_+(\rho_{\rm e},\kappa)$ may be greater than zero at medium densities, where $|dV_{\rm e}/d\rho|$ is large according to empirical observations. 
The related instability mechanism is based on a reduction of the average velocity with increasing density. Due to the continuity equation,
this tends to cause a further compression (but the ``traffic pressure'' terms $P_1$ and $P_2$
partially counteract this re-inforcement mechanism).
\par
As a consequence of the inequality (\ref{conseq}), we can state that
the speed-dependence of the traffic pressure term $P_2$ and the optimal velocity $V_{\rm o}$ tends to make traffic flow more stable with respect to perturbations. The speed-dependence also
resolves problems related to the fact that $\partial P_1/\partial \rho$ may become negative in a certain
density range. This would imply a negative discriminant of the square root, if the negative contribution 
$\partial P_1/\partial \rho < 0$ was
not compensated for by $(\partial P_2/\partial V)^2/(4\rho_{\rm e}{}^2)$ \cite{EPJB}. The case $\partial P_1/\partial \rho <0$ could also cause negative accelerations and speeds, particularly at the end of congestion areas, which would not be realistic \cite{Daganzo}. Again, the second pressure contribution $P_2$ can resolve the problem, if properly chosen.

\subsection{Characteristic Speeds, Phase, and Group Velocities} \label{Macspeeds}

When neglecting the relaxation term (i.e. in the limit $\tau \rightarrow \infty$), the so-called characteristics may be imagined as (parametrized) space-time lines, along which the solution of a macroscopic traffic model based on partial differential equations does not change in time. In Appendix \ref{HPDE}, we derive the characteristics of the linearlized equations (\ref{linrho}) and (\ref{linvau3}). In the following, we will compare the characteristic speeds $C_j(\rho_{\rm e}) = V_{\rm e}(\rho_{\rm e})+c_j(\rho_{\rm e})$ given by Eq. (\ref{charsp}) with the phase velocities $V_{\rm e}(\rho_{\rm e}) + \tilde{\omega}_\pm(\rho_{\rm e},\kappa)/\kappa$ and the group velocities $V_{\rm e}(\rho_{\rm e}) + \partial\tilde{\omega}_\pm(\rho_{\rm e},\kappa)/\partial\kappa$ resulting from the above linear instability analysis. While the phase velocity describes the propagation of a single wave mode, the group velocity describes the propagation of a wave packet composed of waves with different wave numbers $\kappa$ (see Appendix \ref{Groupvel} for details). The group velocity is usually considered to represent the speed of information propagation.\footnote{A typical example is the modulation of electromagnetic waves used to transfer information via radio.} Due to dispersion effects, we may have $\partial\tilde{\omega}_\pm(\rho_{\rm e},\kappa)/\partial\kappa \ne \tilde{\omega}_\pm(\rho_{\rm e},\kappa)/\kappa$.
\par
Let us first study the situation  in the limit $\tau \rightarrow \infty$ of arbitrarily slow adaptation to changed traffic conditions. Considering the definitions (\ref{defi1}) to (\ref{defi3}), we find $1/\hat{\tau}(\kappa) = 0$, $|\Im(\rho_{\rm e},\kappa)| = 0$, and
\begin{equation}
\Re(\rho_{\rm e},\kappa) = - \kappa^2 \frac{\partial  P_1}{\partial \rho} 
- \frac{\kappa^2}{4\rho_{\rm e}{}^2} \left( \frac{\partial P_2}{\partial V}\right)^2 \, .
\end{equation}
For $\Re \le 0$, we have $\sqrt{\Re^2 + \Im^2} = |\Re| = - \Re$ and, due to Eqs. (\ref{real}) and
(\ref{imag}), we obtain
\begin{equation}
\lambda_\pm = 0 \qquad \mbox{and} \qquad \tilde{\omega}_\pm =  - \frac{\kappa}{2\rho_{\rm e}} \left| \frac{\partial P_2}{\partial V}\right| \mp \sqrt{|\Re(\rho_{\rm e},\kappa)|} 
\end{equation}
in the limit $\tau \rightarrow \infty$. This implies 
\begin{eqnarray}
\frac{\partial\tilde{\omega}_\pm(\rho_{\rm e},\kappa)}{\partial\kappa} 
&=& \frac{\tilde{\omega}_\pm(\rho_{\rm e},\kappa)}{\kappa} \nonumber \\
&=&  - \frac{1}{2\rho_{\rm e}} \left| \frac{\partial P_2}{\partial V}\right| 
\mp \sqrt{\frac{\partial  P_1}{\partial \rho} 
+ \frac{1}{4\rho_{\rm e}{}^2} \left( \frac{\partial P_2}{\partial V}\right)^2 } . \qquad
\end{eqnarray}
Therefore, group and phase velocity in the limit $\tau \rightarrow \infty$ are the same. A comparison with Eq. (\ref{charsp}) shows that they also agree with the characteristic speeds. This is expected, because of $\lambda_\pm=0$, which means that the wave amplitudes do not grow or decay---they just propagate along the characteristics.
\par
For {\it finite} values of $\tau$, which are typical for {\it real} traffic flows, the phase and group velocities may be different, and they also do not need to agree with the characteristic speeds, as we will see below: The group velocities, i.e. the propagation speeds of small perturbations, are given by
\begin{eqnarray}
 C_l(\rho_{\rm e},\kappa ) &=& \frac{\partial\omega_l(\rho_{\rm e},\kappa )}{\partial \kappa } 
 = V_{\rm e}(\rho_{\rm e})+ \frac{\partial\tilde{\omega}_l(\rho_{\rm e},\kappa )}{\partial \kappa } \nonumber \\  &=& V_{\rm e}(\rho_{\rm e})+ c_l(\rho_{\rm e},\kappa ) \, ,
\end{eqnarray}
as derived in Appendix \ref{Groupvel}. Obviously, there are two group velocities $C_\pm = V_{\rm e} + c_\pm$, which can be determined by differentiation of the expression for $\tilde{\omega}_\pm(\rho_{\rm e},\kappa)$ given in Eq. (\ref{imag}):
\begin{equation}
c_\pm(\rho_{\rm e},\kappa) = + \frac{1}{2\rho_{\rm e}} \frac{\partial P_2}{\partial V}
\mp \frac{\partial}{\partial \kappa} \sqrt{\frac{1}{2}
 \left( \sqrt{\Re^2 + \Im^2} - \Re \right) } \, .
 \label{cspeed}
\end{equation}
Considering $\partial P_2/\partial V \le 0$ and
\begin{eqnarray}
 \frac{1}{2} \left(\sqrt{\Re^2+\Im^2} - \Re \right) &=& \frac{1}{2} \left(\sqrt{\Re^2+\Im^2} + \Re \right) - \Re 
\nonumber \\ &=& \left( \lambda_\pm + \frac{1}{2\hat{\tau}} \right)^2 - \Re \, ,
\end{eqnarray}
which is implied by Eqs. (\ref{real}) and (\ref{imag}), we may also write 
\begin{equation}
c_\pm (\rho_{\rm e},\kappa) = - \frac{1}{2\rho_{\rm e}} \left| \frac{\partial P_2}{\partial V} \right|
\mp \frac{\partial}{\partial \kappa}
\sqrt{\left( \lambda_\pm + \frac{1}{2\hat{\tau}} \right)^2 - \Re} \, .
\label{reveal}
\end{equation}
Taking into account Eq. (\ref{defi2}), this is generally not the same as $\tilde{\omega}_\pm(\rho_{\rm e},\kappa)/\kappa$, i.e. the phase velocities differ. Interestingly enough, however, at the stability threshold given by $\lambda_{_+} = 0$, we find
\begin{eqnarray}
c_{_+} (\rho_{\rm e},\kappa) &=& - \frac{1}{2\rho_{\rm e}} \left|  \frac{\partial P_2}{\partial V} \right|
- \frac{\partial}{\partial \kappa} \sqrt{\frac{1}{4\hat{\tau}^2}  - \Re} \nonumber \\ 
&=& - \frac{1}{2\rho_{\rm e}} \left| \frac{\partial P_2}{\partial V} \right|- \sqrt{  \frac{\partial  P_1}{\partial \rho} 
+ \frac{1}{4\rho_{\rm e}{}^2} \left( \frac{\partial P_2}{\partial V}\right)^2} \, . \qquad
\end{eqnarray}
At the stability threshold we furthermore have $\lambda_- = -1/\hat{\tau}$. Inserting this into Eq. (\ref{reveal}) reveals
\begin{eqnarray}
c_{_-} (\rho_{\rm e},\kappa) &=& - \frac{1}{2\rho_{\rm e}} \left|\frac{\partial P_2}{\partial V}\right|
+ \frac{\partial}{\partial \kappa}
\sqrt{\frac{1}{4\hat{\tau}^2}  - \Re} \nonumber \\
&=& - \frac{1}{2\rho_{\rm e}} \left| \frac{\partial P_2}{\partial V}\right| + \sqrt{  \frac{\partial  P_1}{\partial \rho} + \frac{1}{4\rho_{\rm e}{}^2} \left( \frac{\partial P_2}{\partial V}\right)^2} \, . \qquad 
\label{AGAIN}
\end{eqnarray}
The same expressions are found for the phase velocities. A comparison with Eq. (\ref{charsp}) shows that they also agree with the characteristic speeds. Note that $c_+$ is {\it smaller} than zero. However, we have $c_- \le 0$ (corresponding to characteristic speeds {\it slower} than the average vehicle velocity or equal to it) {\it only} if 
\begin{equation}
\sqrt{  \frac{\partial  P_1}{\partial \rho} + \frac{1}{4\rho_{\rm e}{}^2} \left( \frac{\partial P_2}{\partial V}\right)^2} \le  \frac{1}{2\rho_{\rm e}} \left| \frac{\partial P_2}{\partial V}\right| 
\end{equation}
or 
\begin{equation}
0 \le - \frac{\partial P_1}{\partial \rho} \le \frac{1}{4\rho_{\rm e}{}^2} \left( \frac{\partial P_2}{\partial V}\right)^2 \, . 
\end{equation}

\section{Discussion} \label{Discuss}

For the discussion of our results regarding the characteristic speeds, let us study two particular models first, the Payne model \cite{Payne1,Payne2} and the Aw-Rascle model \cite{Rascle}.

\subsection{Characteristic Speeds in the Aw-Rascle Model}\label{AR}

The model proposed by Aw and Rascle \cite{Rascle} corresponds to Eqs. (\ref{g1}) and (\ref{veleq}) with $\tau \rightarrow \infty$,
\begin{equation}
\frac{\partial P_1(\rho,V)}{\partial \rho} = 0 \qquad \mbox{and} \qquad \frac{\partial P_2(\rho,V)}{\partial V} = - \gamma \rho(x,t)^{\gamma+1} \le 0 \, ,
\end{equation}
see Ref. \cite{EPJB}. $\gamma$ is a positive constant. This implies $1/\hat{\tau} = 0$, $\Re(\kappa) =  
- \kappa^2(\partial P_2/\partial V)^2/(4\rho_{\rm e}{}^2) < 0$ and $|\Im(\kappa)| = 0$. Therefore,
Eq. (\ref{cspeed}) implies
\begin{eqnarray}
c_\pm(\rho_{\rm e},\kappa) &=& - \frac{1}{2\rho_{\rm e}} \left| \frac{\partial P_2}{\partial V} \right|
\mp \frac{\partial}{\partial \kappa} \sqrt{\frac{1}{2}
 \left( |\Re| - \Re \right) } \nonumber \\
 &=& - \frac{1}{2\rho_{\rm e}} \left| \frac{\partial P_2}{\partial V} \right| \mp \frac{1}{2\rho_{\rm e}} \left| \frac{\partial P_2}{\partial V} \right| \, . 
\end{eqnarray}
This leads to $c_+ = -\gamma \rho(x,t)^{\gamma}$ and $c_- = 0$, corresponding to the characteristic speeds $V -\gamma \rho(x,t)^{\gamma}$ and $V$, in agreement with Aw's and Rascle's calculations \cite{Rascle}. {That is, their model does not have a characteristic speed faster than the average vehicle speed, which elegantly avoids the problem raised by Daganzo \cite{Daganzo}. 
\par
However, is it really necessary to {\it exclude} the existence of a characteristic speed faster than the vehicle speeds? In order to address this problem, we will now study Payne's macroscopic traffic model, which has received most of the criticism. We do this primarily for the sake of illustration, while we are well aware of the weaknesses of this model (like the possibility of backward moving vehicles at upstream jam fronts for certain initial conditions). Therefore, the authors of this paper generally prefer the use of {\it non-local} macroscopic traffic models \cite{EPJB}, but this is not the issue to be discussed, here.}

\subsection{Payne's Traffic Model}

Payne's macroscopic traffic model \cite{Payne1,Payne2} has a solely density-dependent optimal velocity
\begin{equation}
V_{\rm o}(\rho,V)=V_{\rm e}(\rho) \label{Pay1}
\end{equation}
and the pressure gradients 
\begin{equation}
\frac{\partial P_1(\rho,V)}{\partial \rho} = \frac{1}{2\tau} \left| \frac{dV_{\rm e}(\rho)}{d\rho}\right| \ge 0 \,  ,  \qquad \frac{\partial P_2(\rho,V)}{\partial V} = 0\, .\label{Pay2}
\end{equation}
This simplifies the instability condition (\ref{conseq}) considerably, and we get 
\begin{equation}
  \rho_{\rm e} \left| \frac{d V_{\rm e}(\rho_{\rm e})}{d \rho} \right|
 > \frac{1}{2\rho_{\rm e}\tau} \, .
 \label{Payneinstab}
\end{equation}
Traffic flow becomes unstable, if the equilibrium velocity $V_{\rm e}(\rho)$ decreases too rapidly with an increase in the density $\rho$, and greater relaxation times $\tau$ tend to imply larger instability regimes. For the characteristic speeds at the instability threshold, with 
$\rho_{\rm e} | d V_{\rm e}/d \rho | = 1/(2\rho_{\rm e}\tau)$ we find
\begin{equation}
c_\pm (\rho_{\rm e}) 
= \mp \sqrt{  \frac{\partial  P_1}{\partial \rho} }
= \mp \sqrt{ \frac{1}{2\tau} \left| \frac{dV_{\rm e}(\rho_{\rm e})}{d\rho}\right| }
= \mp \rho_{\rm e} \left| \frac{dV_{\rm e}(\rho_{\rm e})}{d\rho}\right| \, .
\label{Paynespeed}
\end{equation}
Clearly, $c_-(\rho)$ is non-negative, i.e. the related characteristic speed 
$V_{\rm e}(\rho) +c_- (\rho)$ tends to be larger than the average vehicle speed $V_{\rm e}(\rho)$.
Nevertheless, by demanding $V_{\rm e}(\rho) +c_-(\rho) \le V^0$, e.g. by assuming a linear 
speed-density function
\begin{equation}
V_{\rm e}(\rho) = V^0\left( 1-\frac{\rho}{\rho_{\rm jam}}\right)\, , \label{demanding}
\end{equation}
one could still reach that the characteristic speed $V_{\rm e}(\rho) +c_-(\rho)$ lies within the variability of the vehicle speeds. In fact, we have $c_\pm = 0$ whenever the vehicle speed cannot vary, namely at density zero and at maximum density, where $\rho_{\rm e}|dV_{\rm e}(\rho_{\rm e})/d\rho| = 0$. However, do we {\it need} to impose such conditions on the characteristic speed and the speed-density relationship? This shall be addressed in the following and in Sec. \ref{Micinstab}.
\par
In connection with this question, it is interesting to note that, according to Eqs. (\ref{AGAIN}) and (\ref{Paynespeed}), the group velocity $c_{_+}$ corresponding to the solution with the {\it unstable} eigenvalue $\lambda_{_+}$ is {\it negative} with respect to the average velocity $V_{\rm e}$. In contrast, propagation at the positive speed $c_{_-}$ with respect to the average velocity $V_{\rm e}$ is related with an {\it eigenmode} that {\it decays quickly, basically at the rate at which the vehicle speeds adjust.} Therefore, the forwardly propagating mode cannot emerge by itself. It could only be produced by a particular specification of the initial condition, enforcing a finite amplitude of the forwardly moving mode.
We will come back to this in Sec. \ref{Micinstab}.
\par 
It is noteworthy that already Whitham performed a thorough analysis of the speeds characterizing the traffic dynamics in what is known as the Payne model today (see Ref. \cite{Whitham}, Chaps. 3 and 10). 
He showed 
that the linearized partial differential equations (\ref{linrho}) and (\ref{linvau3}), when specified in accordance with Eqs. (\ref{Pay1}) and (\ref{Pay2}), can be cast into the equation
\begin{eqnarray}
& & \frac{\partial \delta \rho(x,t)}{\partial t} + \left(V_{\rm e}(\rho) + \rho\frac{dV_{\rm e}(\rho)}{d\rho}\right)
\frac{\partial \delta \rho(x,t)}{\partial x} \nonumber \\
&=& - \tau \left( \frac{\partial}{\partial t} + \big[V_{\rm e}(\rho) + c_+(\rho)\big] \frac{\partial}{\partial x}\right) 
\nonumber \\
& & \times \left( \frac{\partial}{\partial t} + \big[V_{\rm e}(\rho) + c_-(\rho)\big] \frac{\partial}{\partial x}\right) \delta \rho(x,t) \, . \qquad
\end{eqnarray}
Whitham was perfectly aware of the fact that the characteristic speed $V_{\rm e}(\rho) + c_-(\rho)$ was faster than the average vehicle velocity $V_{\rm e}(\rho)$, but not at all worried about this. His perception was that all three velocities were meaningful, and that the kinematic speed $V_{\rm e}(\rho) + \rho\, dV_{\rm e}/d\rho$ would dominate in the limit of small values of $\tau$ (which implies stable vehicle flows).
However, the open problem is still, how a characteristic speed $V_{\rm e}(\rho) + c_-(\rho) > V_{\rm e}(\rho)$ can be interpreted, without violating causality.

\subsection{Characteristic Speeds vs. Vehicle Speeds}\label{versus}

In physical systems, it is not necessarily surprising to find characteristic speeds faster than the average speed. Let us illustrate this for the example of sound propagation. In one spatial dimension, this is described by the continuity equation (\ref{g1})
in combination with the one-dimensional velocity equation
\begin{equation}
 \frac{\partial V(x,t)}{\partial t} + V(x,t) \frac{\partial V(x,t)}{\partial x} = 
 - \frac{1}{\rho} \frac{\partial {\cal P}(\rho)}{\partial x} \, .
\label{g2}
\end{equation}
These so-called Euler equations \cite{hydro1}
can be considered to model frictionless fluid or gas flows in one dimension.
Compared to the velocity equation (\ref{veleq}),
we have dropped the relaxation term $[V_{\rm e}(\rho) - V]/\tau$. Therefore, we do not have an equilibrium velocity-density relation
$V_{\rm e}(\rho)$, now.
\par
In order to determine the solution of the above equations, one can derive linearized equations
for the case of small deviations 
$\delta \rho(x,t) = \rho(x,t) - \rho_{\rm e}$ and $\delta V(x,t) = V(x,t) - V_{\rm e}$ 
from the stationary and homogeneous solution $\rho(x,t) = \rho_{\rm e}$ and $V(x,t) = V_{\rm e} = 0$.
The quantity $\rho_{\rm e}$ corresponds to the average density of the fluid or gas.
\par
Inserting (\ref{Devia}) into Eqs. (\ref{g1}) and (\ref{g2}) and neglecting non-linear terms in 
the small deviations $\delta \rho$, $\delta V$ results in
\begin{equation}
 \frac{\partial \delta \rho(x,t)}{\partial t} + V_{\rm e} \frac{\partial \delta \rho(x,t)}{\partial x}
 = - \rho_{\rm e} \frac{\partial \delta V(x,t)}{\partial x}
\label{g3}  
\end{equation}
and 
\begin{equation}
 \frac{\partial \delta V(x,t)}{\partial t} + V_{\rm e} \frac{\partial \delta V(x,t)}{\partial x} = 
 - \frac{1}{\rho_{\rm e}} \frac{d {\cal P}(\rho_{\rm e})}{d\rho}\frac{\partial \delta \rho(x,t)}{\partial x} \, . 
\label{g4}
\end{equation}
Considering $V_{\rm e} =0$, deriving Eq. (\ref{g3}) with respect to $t$, and Eq. (\ref{g4}) with
respect to $x$ yields 
\begin{equation}
 \frac{\partial^2 \delta \rho(x,t)}{\partial t^2} + \rho_{\rm e} 
\frac{\partial^2 \delta V(x,t)}{\partial t \, \partial x} = 0
\label{g5}  
\end{equation}
and 
\begin{equation}
 \frac{\partial^2 \delta V(x,t)}{\partial x \, \partial t} = 
 - \frac{1}{\rho_{\rm e}} \frac{d {\cal P}(\rho_{\rm e})}{d\rho}\frac{\partial^2 \delta \rho(x,t)}{\partial x^2} \, .
\label{g6}
\end{equation}
Inserting Eq. (\ref{g6}) into Eq. (\ref{g5}) finally gives the so-called wave equation 
\begin{equation}
 \frac{\partial^2 \delta \rho(x,t)}{\partial t^2} - \hat{c}^2 \frac{\partial^2 \delta \rho(x,t)}{\partial x^2}
 = 0 \, ,
\label{sound}
\end{equation}
which is well-known from one-dimensional sound propagation. The constant
\begin{equation}
 \hat{c} = \sqrt{\frac{d{\cal P}(\rho_{\rm e})}{d\rho}} \, ,
 \label{cpress}
\end{equation}
corresponds to the speed of sound. In order to determine the spatio-temporal solution of Eq. (\ref{sound}), we rewrite this equation, inspired by the relationship $(a^2 - b^2) = (a+b)(a-b)$:
\begin{equation}
 \left( \frac{\partial}{\partial t} + \hat{c} \frac{\partial}{\partial x}\right)
 \left( \frac{\partial}{\partial t} - \hat{c} \frac{\partial}{\partial x}\right) \delta \rho(x,t) = 0 \, .
\label{pm}
\end{equation}
According to this equation, perturbations propagate backward and forward at the speed $\pm \hat{c}$,
although the average speed is $V=0$. However, for gases we may assume an approximate pressure law of the form ${\cal P} = \rho \theta_0$ \cite{hydro1}, where $\theta_0$ is the velocity variance of gas molecules. Hence, the speed of sound is given by $\hat{c} = \sqrt{\theta_0}$, i.e. by the standard deviation of velocities. As a consequence, the speed of sound {\it can} actually be propagated by the mobility of gas molecules.
\par
In a similar way, we can understand characteristic speeds faster than the average vehicle speed in the macroscopic model of Phillips \cite{Phillips} or K\"uhne \cite{Kuehne}, Kerner and Konh\"auser \cite{KK}, and Lee {\it et al.} \cite{Lee}. Their pressure functions are also given by the formula ``density times velocity variance''. {Therefore, the faster characteristic speed of {\it these} macroscopic traffic models is expected to lie within the range of individual vehicle speeds.}\footnote{Note that the existence of perturbations in the traffic flow always implies a variation of the vehicle speeds.}
\par
As we have seen above, the situation is generally different for Payne's model. However, it is
illustrative to note that $V_{\rm o}(\rho) + c_{_+}(\rho)$ may become {\it negative}, even when all vehicles move {\it forward}. {That is, it is {\it possible} to have characteristic speeds {\it outside} of the range of vehicle speeds:} According to Eqs. (\ref{Paynespeed}) and (\ref{tilde}), the {slower characteristic speed at the instability threshold is} 
\begin{eqnarray}
& & V_{\rm e}(\rho) + c_{_+}(\rho) = 
V_{\rm e}(\rho) - \rho \left| \frac{d V_{\rm e}(\rho)}{d\rho}\right| \nonumber \\
&=& V_{\rm e}(\rho) + \rho \frac{d V_{\rm e}(\rho)}{d\rho}
= \frac{d Q_{\rm e}(\rho)}{d\rho} \, .
\end{eqnarray}
Since $Q_{\rm e}(\rho) = \rho V_{\rm e}(\rho)$ represents the ``fundamental diagram'', $dQ_{\rm e}(\rho)/d\rho$ describes the negative speed of kinematic waves in the congested regime \cite{Whitham}. 
This does not constitute {\it any} theoretical inconsistency, even if  $V_{\rm e}(\rho_{\rm e}) + c_{_+}(\rho) < 0$. {In fact, we all know situations involving negative group velocities} from dissolving congestion fronts, e.g. when a traffic light turns green: There, the negative propagation speed just results from the fact that the congestion front moves backward, whenever vehicles leave a congested area with some delay. Hence, the negative characteristic speed does not describe the speed of cars. It reflects the propagation of {\it gaps} rather than vehicles. 
\par
{Therefore, could we have a similar mechanism that generates characteristic speeds {\it faster} than the vehicle speeds? If vehicles would react to their leaders with a {\it negative} delay, this would in fact be the case, but it would violate causality. Therefore, all possible explanations for characteristic speeds faster than the vehicle speeds considered so far have failed to resolve the problem. However, the problem may still be a result of the approximations underlying second-order macroscopic traffic models. As we have indicated before, the gradient expansion required to derive them implies some degree of backward interactions. Therefore, it is conceivable that following vehicles would cause their leaders to accelerate, even beyond their desired speed $V^0$. 
\par
If this would be the explanation of a characteristic speed faster than the average speed $V$ or free speed $V^0$, we should not observe it in microscopic traffic models with forward interactions only. Therefore, we will now determine the characteristic speeds of the optimal velocity model \cite{Bando}. This car-following is chosen, because the Payne model can be considered as a macroscopic approximation of it (see \cite{EPJB} and references therein). Besides, we will compare the instability conditions of both models.}

\section{Linear Instability and Characteristic Speeds of the Optimal Velocity Model} \label{Micinstab}

We have seen that macroscopic traffic models behave unstable with respect to small perturbations in a certain density range, where the average velocity changes too rapidly with the density. The same is true for many car-following models. As an example, we will shortly discuss the dynamic behavior of the optimal velocity model. While its stability has been already studied in the past \cite{Bando}, we will focus here on the characteristic speeds, in order to show that characteristic speeds greater than the average velocity are not an artifact of macroscopic traffic models.
\par 
According to the optimal velocity model, the change of the speed $v_i(t)$ of vehicle $i$ is given by
\begin{equation}
\frac{dv_i}{dt} = \frac{v_{\rm o}\big(d_i(t)\big) - v_i(t)}{\tau} 
\label{OptVel}
\end{equation}
and the temporal change of the distance $d_i(t) = x_{i-1}(t) - x_i(t)$ to the leading vehicle $i-1$ is
determined by
\begin{equation}
\frac{dd_i}{dt} = v_{i-1}(t) - v_i(t) \, .
\label{toge}
\end{equation}
In the above equations, the distance-dependent function $v_{\rm o}(d_i)$ is called the optimal velocity function and $\tau$ is again the relaxation time for adjustments of the speed.
\par
\begin{figure}[htbp]
\begin{center}
\includegraphics[width=8.5cm]{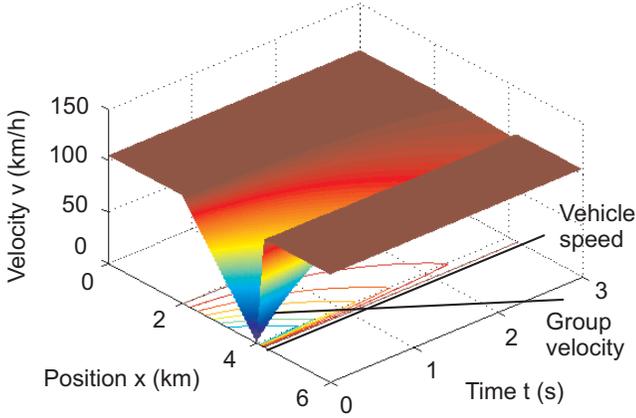}
\end{center}
\caption[]{Simulation result of the optimal velocity model with $v_{\rm o}(d) = v^0\big\{ \tanh[(d-l)/s_0 -1.2] + \tanh(1.2)\big\}/2$, $v^0=115$~km/h, $s_0 = 50$~m, and $l=4$~m.
We have chosen a particular initial condition, where all vehicles started with a distance
$d_{\rm e}=200$~m to their respective leader, but some vehicles $i$ had a speed $v_i(0) < v_{\rm o}(d_{\rm e})$ in the beginning. As a consequence, these vehicles adjusted their speeds to the optimal velocity. The relevant point here is that followers reach the optimal velocity (or certain fractions of it) earlier than their respective leaders. That is, for the particular initial condition chosen here, the perturbation in the speeds propagates {\it faster} than the vehicle speeds. This effect, however, does not violate causality, as the earlier acceleration of upstream cars is not triggered by interactions with followers---it just results from the relaxation term. Therefore, the perturbation disappears on a time scale that is determined by the relaxation time $\tau=1~s$, as predicted by the real part of the eigenvalue $\tilde{\lambda}_-$, see Eq. (\ref{lab}). The relaxation takes longer for larger values of $\tau$. In the limit $\tau\rightarrow \infty$, the perturbation does not decay anymore, but according to Eq. (\ref{lab}), we then have $c_\pm \rightarrow 0$. Therefore, despite its fast speed, the perturbation did not overtake the first car upstream of the initial perturbation in our simulations, when the parameters were chosen in a way that avoided accidents. This confirms the validity of the causality principle.}
\label{FORWARD}
\end{figure}

Appendix \ref{ApPe} sketches the linear stability analysis of the optimal velocity model.
In the following, we will focus on the analysis of the group velocity $c_\pm$ with respect to the average velocity $v_{\rm o}(d_{\rm e})$, i.e. the velocity at which perturbations are expected to propagate.
Relative to the average motion of vehicles with speed $v_{\rm e}(d_{\rm e})$,  
the characteristic speeds are 
\begin{eqnarray}
 c_\pm(d_{\rm e},k) &=& \frac{\partial\tilde{\omega}_\pm(d_{\rm e},\kappa)}{\partial\kappa} 
 = \frac{L}{2\pi} \, \frac{\partial\tilde{\omega}_\pm(d_{\rm e},k)}{\partial k} \nonumber \\
 &=&  \mp \frac{L}{2\pi} \frac{\partial}{\partial k}
 \sqrt{\frac{1}{2} \Big( \sqrt{\Re^2 + \Im^2} - \Re \Big) } \, .
\end{eqnarray}
This can be derived analogously to Eq. (\ref{cspeed}), using Eq. (\ref{magic}) and $\kappa = 2\pi k/L$.
According to Eq. (\ref{reveal}) and due to the series expansion $\cos(x) \approx 1 - x^2/2$, at the instability threshold with $\lambda_+ = 0$ and $dv_{\rm o}(d_{\rm e})/dd = 1/(2\tau) $, we obtain with Eq. (\ref{re})
\begin{eqnarray}
c_\pm (d_{\rm e},k) &=& \mp \frac{L}{2\pi} \frac{\partial}{\partial k} \sqrt{\left( 
\frac{1}{2\tau}\right)^2 - \Re} \nonumber \\
&=&  \mp \frac{L}{2\pi} \frac{\partial}{\partial k} \sqrt{\frac{1}{\tau} \, \frac{dv_{\rm o}(d_{\rm e})}{dd}\big[ 1 - \cos(2\pi k/N)\big] } \nonumber \\
&\approx &  \mp \frac{L}{2\pi} \frac{\partial}{\partial k} \sqrt{ \frac{1}{\tau} \, \frac{dv_{\rm o}(d_{\rm e})}{dd}
\frac{1}{2} \left( \frac{2\pi k}{N}\right)^2 } \nonumber \\
&=& \mp \frac{L}{N} \sqrt{ \frac{1}{2\tau} \, \frac{dv_{\rm o}(d_{\rm e})}{dd} } 
= \mp d_{\rm e} \sqrt{ \frac{1}{2\tau} \, \frac{dv_{\rm o}(d_{\rm e})}{dd} } \label{merk} \\
&=& \mp d_{\rm e}  \sqrt{ \left( \frac{dv_{\rm o}(d_{\rm e})}{dd} \right)^2 } 
= \mp d_{\rm e} \frac{dv_{\rm o}(d_{\rm e})}{dd} \, .
\label{optspeed}
\end{eqnarray}
It is remarkable that the group velocity of the optimal velocity model
can again exceed the average vehicle velocity $ v_{\rm o}(d_{\rm e})$, namely by an amount $c_{_-}(d_{\rm e}) = d_{\rm e} \, dv_{\rm o}(d_{\rm e})/d_{\rm e} > 0$. 
{Moreover, it can be shown that the instability thresholds and the related characteristic speeds 
are the same as for the Payne model (see Appendix \ref{MiMa}). This confirms that the Payne model may be viewed as macroscopic approximation of the optimal velocity model (see \cite{EPJB} and references therein). In view of these results, it is hard to argue that a characteristic speed faster than the vehicle speeds constitutes primarily a theoretical inconsistency of certain kinds of {\it macroscopic} traffic models. Quite unexpectedly, it also occurs for microscopic traffic models that, according to computer simulations, behave reasonably well. 
\par
Therefore, the approximations underlying the Payne model cannot be the problem for the existence of a characteristic speed faster than the vehicle speeds. However, it is interesting to note that the larger group velocity $v_{\rm o}(d_{\rm e}) + c_-(d_{\rm e})$ is related to a {\it negative} real part $\lambda_{_-}$ of the eigenvalue $\tilde{\lambda}_-$. According to Eq. (\ref{cspeed}), the fast characteristic speed $V_{\rm e}(\rho_{\rm e}) + c_-(\rho_{\rm e})$ of macroscopic second-order models is related to a negative eigenvalue $\lambda_-(\rho_{\rm e})$ as well, see Eq. (\ref{real}). Therefore, {\it the related eigenmode decays quickly, and it will be hard to observe in reality. In particular, the faster propagating mode may not emerge by itself.} A closer analysis shows that both, for the optimal velocity model and the Payne model, $\lambda_-$ {\it is of the order $-1/\tau$, i.e. related to the relaxation time $\tau$ of vehicles.} We will see that this observation is highly relevant for understanding perturbations that move faster than the vehicles do.
\par
After all, does the fast characteristic speed {\it really} constitute a theoretical inconsistency? Not so, if we can find initial conditions, for which a following car accelerates or decelerates earlier than the leading car does, although the leader does {\it not} react to the follower. In fact, such initial condition can be constructed:} Figure \ref{FORWARD} shows the result of a computer simulation with $N$ vehicles on a circular road of length $L$. We assume that all vehicles have the distance $d=d_{\rm e}=L/N$ initially. Moreover, all vehicles, with the exception of 10 subsequent vehicles, are assumed to have the initial speed $v_{\rm o}(d_{\rm e})$. Furthermore, the speed of the last of the 10 vehicles is set to 0 (or $v^0$), the speed of the first one to $v_{\rm o}(d_{\rm e})$. The speeds of the vehicles in between are determined by linear interpolation. For this scenario, it is quite natural that the {\it last} of the 10 vehicles accelerates (or decelerates) {\it first}, since it experiences the largest deviation of its actual velocity $v_i(0)$ from the optimal velocity $v_{\rm o}(d_{\rm e})$.  However, {\it as this earlier acceleration (or deceleration) is not interaction-induced, it does not violate causality.} The large characteristic speed in macroscopic traffic models can be understood in a similar way.
 
\section{Summary, Conclusions, and Outlook}\label{Conclusions}

In this paper, we have started with a discussion of Daganzo's sharp criticism of second-order macroscopic traffic flow models \cite{Daganzo}. We have argued that most of the deficiencies identified by Daganzo were fully justified, but could be overcome in the course of time by improved macroscopic traffic models, particularly by non-local multi-class models. However, the issue of characteristic speeds faster than the average vehicle speed was still an open, controversial problem, as it seems to violate causality. In order to study it, we have performed a linear instability analysis of a generalized macroscopic traffic model, which took into account speed-dependencies of the optimal velocity and the traffic pressure terms. Such speed-dependencies occur, for example, in Aw's and Rascle's model \cite{Rascle}. They result when realistic vehicle interactions are considered, and when the possibility of accidents and negative vehicle speeds shall be avoided \cite{IDM,EPJB}. Requirements for reasonable models seem to be
\begin{equation}
\frac{\partial V_{\rm o}(\rho,V)}{\partial \rho} \le 0 \, , \qquad 
\frac{\partial V_{\rm o}(\rho,V)}{\partial V} \le 0\, ,   \qquad 
\frac{\partial P_2(\rho,V)}{\partial V} \le 0 \, ,
\end{equation}
and 
\begin{equation}
\frac{\partial P_1(\rho,V)}{\partial \rho} + \frac{1}{4\rho^2} \left(\frac{\partial P_2(\rho,V)}{\partial V}\right)^2 > 0 \, .
\end{equation}
These conditions are, for example, fulfilled by the gas-kinetic-based traffic model (GKT model), see Ref. \cite{GKTexpans}.
\par
Our main attention was dedicated to the characteristic speeds (or group velocities) rather than the instability thresholds. In the following, we summarize the main results:
\begin{enumerate}
\item {While the characteristic speeds may generally differ from the group and the phase velocities, in the limit $\tau\rightarrow \infty$ of a vanishing source (relaxation) term, they are all the same. Therefore, using a different definition of propagation speeds does not resolve the problem of characteristic speeds faster than the (average or maximum) vehicle speed.}
\item Velocity-dependent pressure terms tend to reduce the characteristic speeds, see Eq. (\ref{reveal}). This is best illustrated by Aw's and Rascle's model, where the fast characteristic agrees with the average vehicle speed.
\item Most macroscopic traffic models have a characteristic speed {\it faster} than the average velocity,
but it may still be within the variability of the vehicle speeds, see Eq. (\ref{demanding}) and Sec. \ref{versus}. 
\item In some models like the Payne model, the characteristic speeds can move slower than the slowest vehicle and faster than the fastest vehicle. The first case is related to delayed acceleration maneuvers at jam fronts and related to gap propagation during jam dissolution, but the second case remained a mystery for a long time. 
\item The {\it faster} characterstic speed is related with a {\it negative} real part of the eigenvalue. This  causes a quick decay of the corresponding eigenmode, basically at the rate, at which the vehicle speed is adjusted. Therefore, this eigenmode will not emerge by itself (see Sec. \ref{Macspeeds}).
\item {If the faster characteristic speed were a result of interactions with following vehicles in a circular road geometry (where {\it following} vehicles influence the {\it downstream} flow as well), the fast eigenmode should decay with the length $L$ of the circular road, not with the relaxation time $\tau$. Therefore, periodic boundary conditions cannot be responsible for a characteristic speed faster than the vehicle speeds. This has also been verified with simulations.}\footnote{Simulations for open boundary conditions basically yield the same results as for periodic boundary conditions, given the system (in terms of the road length $L$) is sufficiently large.}
\item A characteristic speed faster than the vehicle speeds cannot be explained as a result of the approximations underlying macroscopic second-order models, as it is also found for microscopic car-following models, in which vehicle interactions are forwardly directed and velocities are restricted to a range between zero and some maximum speed. For the macroscopic Payne model and the optimal velocity model, we have shown a correspondence not only of the instability thresholds, but also of formulas for the group velocities (see Appendix \ref{MiMa}).
\item Assuming particular initial conditions, characteristic speeds faster than the average vehicle speed could be {\it demonstrated} to exist in computer simulations, where followers accelerate (or decelerate) {\it before} their leaders do (see Fig. \ref{FORWARD}). {\it As these acceleration (or deceleration) processes are induced by artificial initial perturbations rather than by vehicle interactions, this does \underline{\it not} imply a violation of causality.} 
\end{enumerate}
Given these findings, we conclude that characteristic speeds faster than the average speed of vehicles do not constitute a theoretical inconsistency of traffic models and do not need to be ``healed'' by particularly constructed traffic models.\footnote{Of course, this does not speak against models of the Aw-Rascle type.} From our point of view, the problem is that characteristic speeds are hard to imagine. In fact, there is no direct correspondence to particle or vehicle velocities (see Sec. \ref{versus} and Appendix \ref{Groupvel}). The group velocity is nothing more than a matter of phase relations between oscillations of successive vehicles in an eigenmode, and the interpretation as speed of information transmission is sometimes misleading.

\begin{acknowledgement}
{\it Author contributions:} DH performed the analytical calculations and proposed the initial conditions for the simulation presented in Fig. \ref{FORWARD}. AJ generated the computational results and prepared the figure.

{\it Acknowledgment:} DH would like to thank for the inspiring discussions with the participants of the Workshop on ``Multiscale Problems and Models in Traffic Flow'' organized by Michel Rascle and Christian Schmeiser at the Wolfgang Pauli Institute in Vienna from May 5--9, 2008, with partial support by the CNRS.
\end{acknowledgement}

\appendix

\section{Hyperbolic Sets of Partial Differential Equations and Characteristic Speeds}\label{HPDE}

Let us rewrite Eqs. (\ref{linrho}) and (\ref{linvau3}) in the form of a system of linear partial differential equations. With
\begin{eqnarray}
S(\delta \rho,\delta V) &=& \frac{1}{\tau} \bigg[ \frac{\partial V_{\rm o}(\rho_{\rm e},V_{\rm e})}
 {\partial \rho} \, \delta \rho(x,t) \nonumber \\
 & & + \frac{\partial V_{\rm o}(\rho_{\rm e},V_{\rm e})}
 {\partial V} \, \delta V(x,t)  - \delta V(x,t) \bigg]\quad 
\end{eqnarray}
we obtain
\begin{equation}
\frac{\partial}{\partial t} \left(
\begin{array}{c}
\delta \rho(x,t) \\[3mm]
\delta V(x,t)
\end{array} \right) +
\left(\begin{array}{ccc}
 A_{11} & & A_{12} \\[3mm]
 A_{21} & & A_{22} 
\end{array}\right) 
\frac{\partial}{\partial x} \left(
\begin{array}{c}
\delta \rho(x,t) \\[3mm]
\delta V(x,t)
\end{array} \right) = 
\left(
\begin{array}{c}
0 \\[3mm]
S
\end{array} \right)
\label{ReWrite}
\end{equation}
with 
\begin{equation}
\underline{A} = \!
\left(\begin{array}{cc}
 A_{11} & A_{12} \\[3mm]
 A_{21} & A_{22} 
\end{array}\right)\! 
= \!\left(\begin{array}{ccc}
V_{\rm e}(\rho_{\rm e}) & & \rho_{\rm e} \\[3mm]
\frac{1}{\rho_{\rm e}} \frac{\partial P_1(\rho_{\rm e},V_{\rm e})}{\partial \rho} & & 
V_{\rm e}(\rho_{\rm e}) + \frac{1}{\rho_{\rm e}} \frac{\partial P_2(\rho_{\rm e},V_{\rm e})}{\partial V}
\end{array}\right)\! .
\end{equation}
As will be shown below, the solution of this system of partial differential equations
is given by the {\it initial} condition $\delta \rho(x,0)$ and $\delta V(x,0)$. The solution procedure consists basically of two steps: On the one hand, we must determine the so-called characteristics, and on the other hand, we must solve a set of ordinary differential equations to find the solutions along them (see Ref. \cite{NonSci} and footnote 3): With $\vec{u}(x,t)
= \big(\delta \rho(x,t),\delta V(x,t)\big)'$ and $\vec{S} = (0,S)'$ (where the prime indicates a transposed, i.e. a column vector), we can rewrite Eq. (\ref{ReWrite}) as
\begin{equation}
\frac{\partial \vec{u}(x,t)}{\partial t} + \underline{A} \, \frac{\partial \vec{u}(x,t)}{\partial x}
= \vec{S} = \underline{B}\, \vec{u}(x,t) \, . \label{INSERt}
\end{equation}
The source term can be rewritten as $\vec{S}= \underline{B}\,\vec{u}(x,t)$ with
\begin{equation}
\underline{B} = \!
\left(\begin{array}{cc}
 B_{11} & B_{12} \\[3mm]
 B_{21} & B_{22} 
\end{array}\right)\! 
= \!\left(\begin{array}{ccc}
0 & & 0 \\[3mm]
\frac{1}{\tau} \frac{\partial V_{\rm o}(\rho_{\rm e},V_{\rm e})}
 {\partial \rho} & & \frac{1}{\tau}\left( \frac{\partial V_{\rm o}(\rho_{\rm e},V_{\rm e})}
 {\partial V} - 1 \right)  
\end{array}\right)\! .
\end{equation}
Now, let $C_j$ denote the eigenvalues of the matrix $\underline{A}$. The values of $C_j = V_{\rm e}(\rho_{\rm e}) + c_j$ satisfying det$(\underline{A}-C_j\underline{1})=0$ are given by the characteristic polynomial
\begin{equation}
 c_j{}^2 - \frac{c_j}{\rho_{\rm e}}\frac{\partial P_2}{\partial V} - \frac{\partial P_1}{\partial \rho} = 0 \, , 
\end{equation}
which results in 
\begin{equation}
 c_j = \frac{1}{2\rho_{\rm e}} \frac{\partial P_2}{\partial V} \pm 
 \sqrt{ \frac{1}{4\rho_{\rm e}{}^2}\left( \frac{\partial P_2}{\partial V}\right)^2 + \frac{\partial P_1}{\partial \rho} } \, . \label{charsp}
\end{equation}
Furthermore, let $\vec{z}_j$ be the eigenvectors related with the eigenvalues $C_j = V_{\rm e} + c_j$, i.e.
\begin{equation}
 \underline{A} \, \vec{z}_j = C_j \vec{z}_j \, . 
\end{equation}
Finally, let $\underline{R}=(R_{ij})$ be the matrix containing the eigenvectors $\vec{z}_j$ as their $j$th
column, and $\vec{y}(x,t) = \underline{R}^{-1} \vec{u}(x,t)$ or $\vec{u}(x,t) = \underline{R}\, \vec{y}(x,t)$. Then, inserting this into Eq. (\ref{INSERt}) and multiplying the result with the inverse matrix $\underline{R}^{-1}$ of
$\underline{R}$ yields 
\begin{equation}
\frac{\partial y_j(x,t)}{\partial t} + C_j \frac{\partial y_j(x,t)}{\partial x}
= (\underline{R}^{-1}\vec{S})_j = (\underline{R}^{-1} \underline{B}\, \underline{R} \, \vec{y})_j\, .
\label{REWRI}
\end{equation}
For $\vec{S} = 0$ (corresponding to the limiting case $\tau\rightarrow \infty$), we have
\begin{equation}
y_j(x,t) = y_j(x-C_j t,0) \, , \label{NOT1}
\end{equation}
which means that the solution does not change in time along the characteristics
$x_j(t)=C_j t$. The quantities $C_j$ are called the characteristic speeds.\footnote{The
idea behind the characteristics is to introduce a parameterization $t(s_1,s_2)$, $x(s_1,s_2)$, which is defined by $\partial t/\partial s_j = 1$ and $\partial x/\partial s_j = C_j$. Then, one can rewrite Eq. (\ref{REWRI}) as
\[
\frac{\partial y_j}{\partial s_j} = \frac{\partial y_j(x,t)}{\partial t}\, \frac{\partial t}{\partial s_j} 
+ \frac{\partial y_j(x,t)}{\partial x}\, \frac{\partial x}{\partial s_j}  
= (\underline{R}^{-1} \underline{B}\, \underline{R} \, \vec{y})_j \, .
\]
In the generalized coordinates $s_1$ and $s_2$, the partial differential equations in $x$ and $t$ we were starting with, turn into {\it ordinary} differential equations. These are much easier to solve.} 
If $\vec{u}(x,0)$ is the initial condition, the solution of the set of partial differential equations is 
\begin{equation}
u_i(x,t) = \sum_j R_{ij}  y_j(x-C_j t,0) \label{NOT2}
\end{equation}
with $\vec{y}(x,0) = \underline{R}^{-1}\vec{u}(x,0)$.\footnote{Note that formulas (\ref{NOT1}) and (\ref{NOT2}) only apply to the limiting case $\tau\rightarrow \infty$, where the relaxation term of the macroscopic traffic model vanishes.} Therefore, the spatio-temporal solution $\vec{u}(x,t)$ is fully determined by the initial condition. In other words, the future state of the system is given by its previous state, and the principle of causality should be valid.

\section{Stability Analysis for Macroscopic Traffic Models}\label{MAC}

In order to understand the dynamics of traffic flows, it is important to find out whether and under what conditions variations in the traffic flow can grow and eventually cause traffic congestion. 
For this, it is useful to make the solution ansatz
\begin{eqnarray}
 \delta \rho(x,t) &=& \delta \rho_0 \, \exp \big( {\rm i}\kappa  x +  (\lambda - {\rm i} \omega)t \big) 
 = \delta \rho_0 \, \mbox{e}^{\lambda t} \, \mbox{e}^{{\rm i}(\kappa x - \omega t)} \, , 
\nonumber \\ 
 \delta V(x,t) &=& \delta V_0 \, \exp \big( {\rm i}\kappa  x +  (\lambda - {\rm i} \omega)t \big) 
 = \delta V_0 \, \mbox{e}^{\lambda t} \, \mbox{e}^{{\rm i}(\kappa x - \omega t)} \, . \nonumber \\
& & \label{ans}
\end{eqnarray}
Because of $\exp({\rm i}\kappa x) = \cos(\kappa x) + {\rm i} \sin(\kappa x)$ (see Appendix \ref{Magic}), ansatz (\ref{ans}) assumes that the perturbation of the stationary and homogeneous traffic situation can be represented as a periodic function with the wave number $\kappa $ and wavelength $2\pi/\kappa $. The wave frequency of Eq. (\ref{ans}) is $\omega$, while $\delta\rho_0\,\exp(\lambda t)$ and $\delta V_0\,\exp(\lambda t)$ are the amplitudes at time $t$. That is, if the ``growth rate'' $\lambda$ is greater than zero, even small perturbations will 
eventually grow, which can give rise to ``phantom traffic jams''. For $\lambda < 0$, however, the initial perturbation will be damped out and the stationary and homogeneous solutions will be re-established, i.e. it is stable with respect to small perturbations. 
\par
Below we will see that, 
for each specification of $\kappa $ and the average density $\rho_{\rm e}$,
there exist two solutions $l \in \{ +, -\}$ with the frequencies $\omega_l(\kappa )$ and the growth rates 
$\lambda_l(\kappa )$. All the corresponding specifications of ansatz (\ref{ans}) are solutions 
of the linearized partial differential equations. The same applies to their superpositions. 
The general solution for an {\em arbitrary} initial perturbation is of the form 
\begin{eqnarray}
 \delta \rho(x,t) &=& \!\!\!\!\! \sum_{l\in\{+,-\}} \!\!
 \int \!\! d\kappa \, \delta \rho_0^l(\kappa ) \exp\big( {\rm i}\kappa  x + 
 \big[\lambda_l(\kappa ) - {\rm i} \omega_l(\kappa  )\big]t\big) \, ,
\nonumber \\
 \delta V(x,t) &=& \!\!\!\!\! \sum_{l\in\{+,-\}} \!\!
 \int \!\! d\kappa  \, \delta V_0^l(\kappa)  \exp\big( {\rm i}\kappa x + 
 \big[\lambda_l(\kappa ) - {\rm i} \omega_l(\kappa  )\big]t\big) \, .\nonumber \\
 & & \label{four} 
\end{eqnarray}
In order to find the possible $\kappa$-dependent wave numbers $\omega$ and growth rates $\lambda$,
we insert ansatz (\ref{ans}) into the linearized macroscopic traffic equations
(\ref{linrho}) and (\ref{linvau3}) and use the relationship ${\rm i}^2 = -1$. The result can represented as an eigenvalue problem: 
\begin{equation}
\left(\begin{array}{ccc}
 M_{11} & & M_{12} \\[3mm]
 M_{21} & & M_{22} 
\end{array}\right) 
\left(
\begin{array}{c}
\delta \rho_0 \\[3mm]
\delta V_0
\end{array} \right)
\stackrel{!}{=} \left( \begin{array}{c}
0 \\
0
\end{array} \right) \, ,
\label{eigen}
\end{equation}
where
\begin{eqnarray}
M_{11} &=& -\tilde{\lambda}  \, , \\
M_{12} &=& - {\rm i} \kappa  \rho_{\rm e} \, , \\
M_{21} &=& -  \frac{{\rm i} \kappa}{\rho_{\rm e}} \frac{\partial P_1}{\partial \rho}
+ \frac{1}{\tau} \frac{\partial V_{\rm o}}{d\rho} \, , \\
M_{22} &=& - \tilde{\lambda}  - \frac{{\rm i}\kappa}{\rho_{\rm e}} \frac{\partial P_2}{\partial V}
+ \frac{1}{\tau} \frac{\partial V_{\rm o}}{\partial V} - \frac{1}{\tau} 
\end{eqnarray}
and 
\begin{equation}
 \tilde{\lambda} = \lambda - {\rm i} \tilde{\omega} \qquad \mbox{with} \qquad
 \tilde{\omega} = \omega - \kappa  V_{\rm e}(\rho_{\rm e}) \, . 
\end{equation}
Equation (\ref{eigen}) is fulfilled only for certain values of
$\tilde{\lambda}(\kappa )$, the so-called ``eigenvalues''. These depend on the average density $\rho_{\rm e}$ and solve the {\em characteristic polynomial} of second order in $\tilde{\lambda}$, which  
is obtained by determining the determinant 
\begin{equation}
 \mbox{det}(\underline{M}) = M_{11} M_{22} - M_{21}M_{12} 
\end{equation}
of the matrix $\underline{M}$ and 
requiring that it becomes zero. The corresponding characteristic polynomial is given by Eq. (\ref{monopol3}).

\section{Derivation of Formula (19)} \label{Magic}

Remember that a complex number
\begin{equation}
 z = \Re + {\rm i}\Im = r \mbox{e}^{{\rm i}\varphi} = r \cos(\varphi) + {\rm i} r \sin(\varphi) 
 \label{decomp}
\end{equation}
can be represented in two-dimensional space with coordinates $\Re = \mbox{Re}(z) = r \cos(\varphi)$ and $\Im=\mbox{Im}(z) = r\sin(\varphi)$, respectively, called the real part and the imaginary part. The absolute value is given as
\begin{equation}
 r = \sqrt{\Re^2 + \Im^2} = \sqrt{(\Re+ {\rm i}\Im) (\Re - {\rm i}\Im)} = \sqrt{z \, \overline{z}} = |z| \, ,
\end{equation}
where $\overline{z} = \Re - {\rm i} \Im = r \mbox{e}^{-{\rm i}\varphi}$ is the conjugate complex number. 
The angle $\varphi$ is determined by
\begin{equation}
 \tan (\varphi) = \frac{\sin(\varphi)}{\cos(\varphi)} = \frac{\Im}{\Re}  = \frac{\mbox{Im}(z)}{\mbox{Re}(z)} \, ,
\end{equation}
and the exponential functions is defined as for real numbers by the infinite series expansion
\begin{equation}
 \exp(z) = \mbox{e}^{z} = \sum_{l=0}^\infty \frac{z^l}{l!} \, ,
\end{equation} 
where $l! = l \cdot (l-1) \dots 2 \cdot 1$. Therefore, the relationships for exponential functions apply also to the case of complex numbers, i.e. the product of two complex numbers $z_1 = \Re_1 + {\rm i}\Im_1=r_1\mbox{e}^{{\rm i}\varphi_1}$ and
$z_2 = \Re_2 + {\rm i} \Im_2 = r_2 \mbox{e}^{{\rm i}\varphi_2}$ is given by
\begin{eqnarray}
z_1 z_2 &=& \big(\Re_1\Re_2 - \Im_1\Im_2\big) + {\rm i} \big( \Re_1 \Im_2 + \Im_1 \Re_2 \big) \nonumber \\ 
&=& r_1 \mbox{e}^{{\rm i}\varphi_1}r_2 \mbox{e}^{{\rm i}\varphi_2} 
= r_1r_2 \mbox{e}^{{\rm i}(\varphi_1+\varphi_2)}  \nonumber \\
&=& r_1r_2 \cos(\varphi_1+\varphi_2)  + {\rm i} r_1r_2 \sin(\varphi_1+\varphi_2) \, .
\end{eqnarray}
As the real and imaginary part are linearly independent of each other, this implies
$\Re_1\Re_2 - \Im_1\Im_2= r_1r_2 \cos(\varphi_1+\varphi_2) $ and $\Re_1 \Im_2 + \Im_1 \Re_2=r_1r_2 \sin(\varphi_1+\varphi_2)$.
The inverse of a complex number is given by
\begin{equation}
 \frac{1}{z} = \frac{1}{r \mbox{e}^{{\rm i}\varphi}} = \frac{\mbox{e}^{-{\rm i}\varphi}}{r} \, .
\end{equation}
The imaginary unit ${\rm i}$ has the property ${\rm i}^2 = -1$ and may, therefore, be written as
${\rm i} = \sqrt{-1}=\mbox{e}^{{\rm i}\pi/2}$.
\par
The square of complex numbers
\begin{equation}
z = r \mbox{e}^{\pm {\rm i}\varphi} = r \big[\cos(\varphi) \pm {\rm i}\sin(\varphi)\big] \, ,
\end{equation}
can, on the one hand, be written as
\begin{equation}
z^2 = r^2 \Big[ \cos^2(\varphi) \pm 2{\rm i} \cos(\varphi)\sin(\varphi) - \sin^2 (\varphi) \Big] \, .
\end{equation}
On the other hand, using the well-known law $\mbox{e}^{x_1}\cdot \mbox{e}^{x_2} = \mbox{e}^{x_1+x_2}$ for the exponential function, we find the alternative representation
\begin{equation}
z^2 = r^2 \big(\mbox{e}^{\pm {\rm i}\varphi}\big)^2 
 = r^2 \mbox{e}^{\pm {\rm i}2\varphi} = r^2
 \big[\cos(2\varphi) \pm {\rm i}\sin(2\varphi)] \, .
\end{equation} 
Comparing the real parts and using the trigonometric relationship $\sin^2(x)+\cos^2(x) = 1$,
we find
\begin{equation}
\cos(2\varphi) = 1 - 2 \sin^2 (\varphi) = 1 - 2 \big[1-\cos^2(\varphi)\big] = 2\cos^2(\varphi) - 1 \, ,  
\end{equation}
from which we can derive the trigonmetric formulas
\begin{equation}
 \sin^2 (\varphi/2) = \frac{1}{2} \big[1-\cos(\varphi)\big] 
 \end{equation}
 and
 \begin{equation}
 \cos^2(\varphi/2) = \frac{1}{2} \big[1+\cos(\varphi)\big] \, .
\end{equation}
Therefore, the square root of a complex number is given by
\begin{eqnarray}
\sqrt{z} &=& \sqrt{r} \mbox{e}^{\pm {\rm i}\varphi/2} = \sqrt{r} \big[\cos(\varphi/2) \pm {\rm i}\sin(\varphi/2)\big]
\nonumber \\
&=& \sqrt{\frac{1}{2} \big[ r + r \cos(\varphi) \big] } 
\pm {\rm i} \sqrt{\frac{1}{2}  \big[ r - r \cos(\varphi) \big] } \, .
\end{eqnarray}
Considering $\Re = r\cos(\varphi)$, $\Im = r\sin(\varphi)$, and
$\Re^2 + \Im^2 = r^2$, we end up with the
desired equation
\begin{equation}
 \sqrt{\Re \pm {\rm i}|\Im |} = \!\sqrt{\frac{1}{2} \Big( \!\sqrt{\Re^2 + \Im^2} + \Re \Big) }
 \pm {\rm i}  \sqrt{\frac{1}{2} \Big( \! \sqrt{\Re^2 + \Im^2} - \Re \Big) } .
\end{equation}

\section{Meaning of the Group Velocity} \label{Groupvel}

Let us start with the representation (\ref{four}) of the general solution of the linearized system of equations,
focussing (for simplicity) on the case $\lambda_l(\kappa) = 0$ and assuming a ``Gaussian wave packet'' with 
\begin{equation}
\delta \rho_0^l(\kappa ) = \frac{\mbox{e}^{-(\kappa-\kappa_0)^2/(2\theta)}}{\sqrt{2\pi\theta}}   \, .
\label{wpack}
\end{equation}
Via the linear Taylor approximation $\omega_l(\kappa) = \omega_l(\kappa_0) + C_l \, \Delta \kappa$ 
with $C_l = d\omega_l(\kappa_0)/d\kappa$ and $\Delta \kappa = (\kappa - \kappa_0)$, from Eq. (\ref{four}) we get
\begin{eqnarray}
& &  \delta \rho(x,t) \nonumber \\
&=& \sum_{l\in\{+,-\}} \int\limits_{-\infty}^\infty d\kappa \, \frac{\mbox{e}^{-(\kappa-\kappa_0)^2/(2\theta)}}{\sqrt{2\pi\theta}} \mbox{e}^{{\rm i} [ \kappa x - \omega_l(\kappa) t ]} \nonumber \\
 &=& \sum_{l\in\{+,-\}} \mbox{e}^{{\rm i} [ \kappa_0 x - \omega_l(\kappa_0) t ]} 
\int\limits_{-\infty}^\infty d \Delta \kappa \, \frac{\mbox{e}^{-(\Delta \kappa)^2/(2\theta)}}{\sqrt{2\pi\theta}} \mbox{e}^{{\rm i} [ \Delta \kappa x - C_l t ]} \nonumber \\
 &=& \sum_{l\in\{+,-\}} \mbox{e}^{{\rm i} [ \kappa_0 x - \omega_l(\kappa_0) t ]} 
\underbrace{\int\limits_{-\infty}^\infty d \Delta \kappa \, \frac{\mbox{e}^{-[ \Delta \kappa -{\rm i} \theta (x-C_lt) ]^2/(2\theta)}} {\sqrt{2\pi\theta}}}_{=1} \nonumber \\[2mm]
& & \qquad\quad  \times \, \mbox{e}^{- \theta (x-C_lt)^2/2} \nonumber \\[2mm]
 &=& \sum_{l\in\{+,-\}}  \mbox{e}^{{\rm i} [ \kappa_0 x - \omega_l(\kappa_0) t ]} \mbox{e}^{- \theta (x-C_lt)^2/2} \, .
 \label{wpack1}
\end{eqnarray} 
While the single waves of frequency $\omega_l(\kappa)$ move with the ``phase velocity'' $x/t = \omega_l(\kappa)/\kappa$, it turns out that their superposition behaves like a wave with frequency $\omega_l(\kappa_0)$ and speed $x/t = \omega_l(\kappa_0)/\kappa_0$. However, the wave packet or, more exactly speaking, its amplitude $\mbox{e}^{- \theta (x-C_lt)^2/2}$ is
moving with the group velocity $x/t = C_l = d\omega_l(\kappa)/d\kappa$. Note that the case $C_l > \omega_l(\kappa_0)/\kappa_0$, in which the group velocity is greater than the phase velocity (wave velocity), is possible. It is called ``anomalous dispersion''.

\section{Linear Stability Analysis of the Optimal Velocity Model} \label{ApPe}

For a linear stability analysis of the optimal velocity model, we imagine the situation of $N$ vehicles $i$ distributed over a circular road of length $L$. This allows us to assume periodic boundary
conditions. The stationary solution for this case is given by
$dv_i/dt = 0$ and $dd_i/dt = 0$, which implies
\begin{eqnarray}
 d_i(t) &=& d_{\rm e} = L/N = \mbox{const.}  \nonumber \\
 v_{i-1}(t) = v_i(t) &=& v_{\rm o}(d_{\rm e}) = \mbox{const.} 
\end{eqnarray}
We are now interested how the deviations from this solution, i.e. the variables
\begin{eqnarray}
 \delta d_i(t) &=& d_i(t) - d_{\rm e} \, , \nonumber \\
 \delta v_i(t) &=& v_i(t) - v_{\rm o}(d_{\rm e}) \, , 
\end{eqnarray}
develop in time, assuming that the initial deviations are small, i.e. $\delta d_i(0) \ll d_{\rm e}$ and
$\delta v_i(0) \ll v_{\rm e}(d_{\rm e})$. For this, we linearize the model equations (\ref{OptVel}) and
(\ref{toge}) around the stationary and homogeneous solution. This results in
\begin{eqnarray}
 \frac{d\delta v_i(t)}{dt} &=& \frac{1}{\tau} \left( \frac{dv_{\rm o}(d_{\rm e})}{dd} \delta d_i(t)
 - \delta v_i(t) \right) \, , \nonumber \\
 \frac{d\delta d_i(t)}{dt} &=& \delta v_{i-1}(t) - \delta v_i(t) \, .
\label{ins}
\end{eqnarray}
For the analysis of stability, we use the solution ansatz
\begin{eqnarray}
 \delta v_j(t) &=& \delta v_0 \, \mbox{e}^{{\rm i}2\pi jk/N + \tilde{\lambda} t} 
 = \delta v_0 \, \mbox{e}^{{\rm i}j\kappa L/N + \tilde{\lambda} t} \, , \nonumber \\
 \delta d_j(t) &=& \delta d_0 \, \mbox{e}^{{\rm i}2\pi jk/N + \tilde{\lambda} t} 
  = \delta d_0 \, \mbox{e}^{{\rm i} j\kappa L/N + \tilde{\lambda} t} \, ,
 \label{ins1}
\end{eqnarray}
where $\kappa = 2\pi k/L$ is the so-called wave number, which
is inversely proportional to the wave length $2\pi/\kappa = L/k$. 
Note that, due to the assumed periodic boundary conditions, possible wavelength are fractions $L/k$ 
of the length $L$ or the circular road. The shortest wave length is given by the average vehicle distance
$d_{\rm e} = L/N$, i.e. $k \in \{1,2,\dots, N\}$. Summing up the functions (\ref{ins1}) over these values
of $k$ results in the {\em Fourier representation} of $\delta v_j(t)$ and $\delta d_j(t)$:
\begin{eqnarray}
\delta v_j(t) &=& \sum_{k=1}^N \delta v_k \mbox{e}^{{\rm i}2\pi jk/N + \tilde{\lambda} t} \, , \nonumber \\
\delta d_j(t) &=& \sum_{k=1}^N \delta d_k \mbox{e}^{{\rm i}2\pi jk/N + \tilde{\lambda} t} \, .
\end{eqnarray}
The parameters $\delta v_k$ and $\delta d_k$ are determined by the initial conditions of
all vehicles $j$. $\tilde{\lambda} = \lambda- {\rm i} \tilde{\omega}$ 
are the so-called {\em eigenvalues}, whose real part $\lambda$
describes an exponential growth (if $\lambda > 0$) or decay (if $\lambda <0$), and
whose imaginary part $\tilde{\omega}$ reflects oscillation frequencies.
$\delta d_0$ and $\delta v_0$ denote oscillation
amplitudes. Inserting this into (\ref{ins}) and dividing by $\mbox{e}^{{\rm i}2\pi jk/N + \tilde{\lambda} t}$,
we finally obtain
\begin{eqnarray}
 \tilde{\lambda} \delta v_0 &=& \frac{1}{\tau} \left( \frac{dv_{\rm o}(d_{\rm e})}{dd} \delta d_0 -\delta v_0\right) \, , 
\label{f1} \\ 
 \tilde{\lambda} \delta d_0 &=& \delta v_0 \mbox{e}^{-{\rm i}2\pi k/N} - \delta v_0 
 = \delta v_0 \Big(\mbox{e}^{-{\rm i}2\pi k/N} - 1\Big) \, . \qquad 
\label{f2}
\end{eqnarray}
Multiplying Eq. (\ref{f1}) with $\tilde{\lambda}$ and inserting Eq. (\ref{f2}) for $\tilde{\lambda} \, \delta d_0$ in the square brackets gives, after
division by $\delta v_0$, the {\em characteristic polynomial} in the eigenvalues $\tilde{\lambda}$,
namely
\begin{equation}
\tilde{\lambda}^2 + \frac{1}{\tau} \tilde{\lambda} - \frac{1}{\tau}  \frac{dv_{\rm o}(d_{\rm e})}{dd} 
\Big(\mbox{e}^{-{\rm i}2\pi k/N} - 1\Big) = 0\, .
\end{equation}
The solutions $\tilde{\lambda}(d_{\rm e},k)$ of this polynomial are the eigenvalues. They read
\begin{equation}
  \tilde{\lambda}_\pm(d_{\rm e},k) = - \frac{1}{2\tau} \pm \sqrt{ \frac{1}{4\tau^2} + \frac{1}{\tau} \frac{dv_{\rm o}(d_{\rm e})}{dd}
\Big(\mbox{e}^{-{\rm i}2\pi k/N} - 1\Big) } \, .
\label{lab}
\end{equation}
Again, the square root contains a complex number, which makes it difficult to see
the sign of the real value $\lambda_\pm$ of $\tilde{\lambda}_\pm$. However, considering
$\mbox{e}^{\pm {\rm i}\varphi} = \cos(\varphi) \pm {\rm i}\sin(\varphi)$ and defining the
real part 
\begin{equation}
 \Re = \frac{1}{4\tau^2} - \frac{1}{\tau} \frac{dv_{\rm o}(d_{\rm e})}{dd} \big[ 1 - \cos(2\pi k/N)\big]
\label{re}
\end{equation}
of the expression under the root and its imaginary part
\begin{equation}
 \Im = - \frac{\sin(2\pi k/N)}{\tau}  \frac{dv_{\rm o}(d_{\rm e})}{dd} \, ,
\label{im}
\end{equation}
we can again apply the useful formula (\ref{magic}). From this we can conclude that
$\lambda = \mbox{Re} (\tilde{\lambda}) = 0$ if
\begin{equation}
 \frac{1}{16\tau^4} = \frac{\Re}{4\tau^2} + \frac{\Im^2}{4} \, ,
\end{equation}
see Eq. (\ref{sechzehn}). Inserting Eqs. (\ref{re}) and (\ref{im}), we find 
\begin{equation}
 \frac{\sin^2(2\pi k/N)}{4\tau^2} \left(\frac{dv_{\rm o}(d)}{dd}\right)^2 = \frac{1}{4\tau^3} \frac{dv_{\rm o}(d)}{dd} \big[ 1 - \cos(2\pi k/N) \big] \, ,
\end{equation}
which finally results in the condition
\begin{equation}
 \frac{dv_{\rm o}(d_{\rm e})}{dd} = \frac{1-\cos(2\pi k/N)}{\tau \sin^2(2\pi k/N)} \stackrel{k\rightarrow 0}{=}
 \frac{1}{2\tau} \, .
\end{equation}
The limit $2\pi k/N \rightarrow 0$  follows from
$\cos(\varphi) \approx 1-\varphi^2/2$ and $\sin(\varphi) \approx \varphi$ in the limit of small wave numbers $\kappa = 2\pi k/L$, i.e. large wave lengths $2\pi/\kappa = L/k$. 
\par
It can be demonstrated by numerical analyses that
\begin{equation}
 \frac{dv_{\rm o}(d_{\rm e})}{dd} > \frac{1}{2\tau} 
 \label{optinstab}
\end{equation}
constitutes the instability condition of the optimal velocity model (\ref{OptVel}) \cite{Bando}. 
In other words, if the velocity changes too strongly with
the distance, small variations of the vehicle distance or speed will grow and finally cause 
emergent waves, i.e. the formation of one or several traffic jams. Since the origin of such 
a breakdown can be infinitesimally small, these traffic jams seem to have no origin. In such
situations, one speaks of ``phantom traffic jams''. A closer analysis for realistic speed-distance
relationships $v_{\rm o}(d)$ shows that traffic tends to be unstable at medium densities
$\rho =1/d$, while it tends to be stable at small and large densities (where the speed
does not change much with a variation in the distance). Only a sufficient reduction in the adaptation time
$\tau$ can avoid an instability of traffic flow, while large delays in the velocity adjustment lead to
growing perturbations of traffic flow.

\section{Correspondence of the Optimal Velocity Model with the Macroscopic Payne Model}\label{MiMa}

As the Payne model has been claimed to be a macroscopic approximation of the optimal velocity model (see Ref. \cite{EPJB} and citations therein), it is interesting to compare the instability conditions and characteristic speeds of both models. Therefore, let us make the identifications
\begin{equation}
 \rho = \frac{1}{d} \, , \quad V_{\rm e}(\rho) = v_{\rm o}\left(\frac{1}{\rho}\right)\, . 
\end{equation}
Then, with the chain rule and the quotient rule of Calculus we can derive
\begin{eqnarray}
\left|\frac{dV_{\rm e}(\rho)}{d\rho}\right| &=& - \frac{dV_{\rm e}(\rho)}{d\rho}
= -\frac{dv_{\rm o}(1/\rho)}{d\rho} = - \frac{dv_{\rm o}(d)}{dd} \frac{dd}{d\rho} \nonumber \\
&=&   \frac{dv_{\rm o}(d)}{dd} \cdot \frac{1}{\rho^2} \, .
\end{eqnarray}
Inserting this into Eq. (\ref{Payneinstab}) gives
\begin{equation}
  \rho_{\rm e} \left| \frac{d V_{\rm e}}{d \rho} \right|
  =  \frac{1}{\rho_{\rm e}} \frac{dv_{\rm o}(d)}{dd}  
 > \frac{1}{2\rho_{\rm e}\tau} 
\end{equation}
or
\begin{equation} 
  \frac{dv_{\rm o}(d_{\rm e})}{dd} > \frac{1}{2\tau}  \qquad \mbox{and} \qquad
  \rho_{\rm e} \left| \frac{dV_{\rm e}(\rho_{\rm e})}{d\rho}\right| = d_{\rm e} \frac{dv_{\rm o}(d_{\rm e})}{dd} \, ,
\end{equation}
where $d_{\rm e} = 1/\rho_{\rm e}$. This shows the agreement of the instability conditions (\ref{Payneinstab}) and (\ref{optinstab}) and of the characteristic speeds (\ref{Paynespeed}) and (\ref{optspeed}) at the instability threshold.

\end{document}